\documentclass[aps,prb,twocolumn,floatfix,groupedaddress,showkeys]{revtex4}
\usepackage{latexsym} \usepackage{amsmath} \usepackage{dcolumn}
\usepackage{xcolor}
\usepackage{xfrac}
\usepackage{bm} \usepackage{graphicx} \usepackage{epsfig}
\graphicspath{{./Figures/}}

\begin{document}


\title{Non-resonant two-photon x-ray absorption in Cu}




\author{J. J. Kas} \affiliation{Dept.\ of Physics, Univ.\ of
Washington Seattle, WA 98195}
\author{J. J. Rehr} 
\email[]{jjr@uw.edu}
\affiliation{Dept.\ of Physics, Univ.\ of
Washington Seattle, WA 98195}

\author{J. St\"ohr}
 \affiliation{SLAC National Accelerator Laboratory, Menlo Park, CA 94025 }
 \author{J. Vinson}
 \affiliation{Material Measurement Laboratory, National Institute of Standards and Technology, Gaithersburg, Maryland, 20899}

\date{\today}

\begin{abstract}

 We present a real-space Green's function theory and  calculations of
two-photon x-ray absorption (TPA).  Our focus is on non-resonant $K$-shell TPA in metallic Cu, which has been observed experimentally at intense x-ray free electron laser (XFEL) sources.  The theory is based on an independent-particle Green's function treatment of the Kramers-Heisenberg equation and an approximation for the sum over non-resonant intermediate 
states in terms of a static quadrupole transition operator. XFEL effects are modeled by a partially depleted $d$-band.  This approach is shown to give results for $K$-shell TPA in quantitative agreement with  XFEL experiment and with a  Bethe-Salpeter Equation approach.
We also briefly discuss many-body corrections  and TPA sum-rules.  

\end{abstract}


\maketitle

\section{Introduction}

\begin{figure}[b]
\includegraphics[width=0.6\columnwidth]{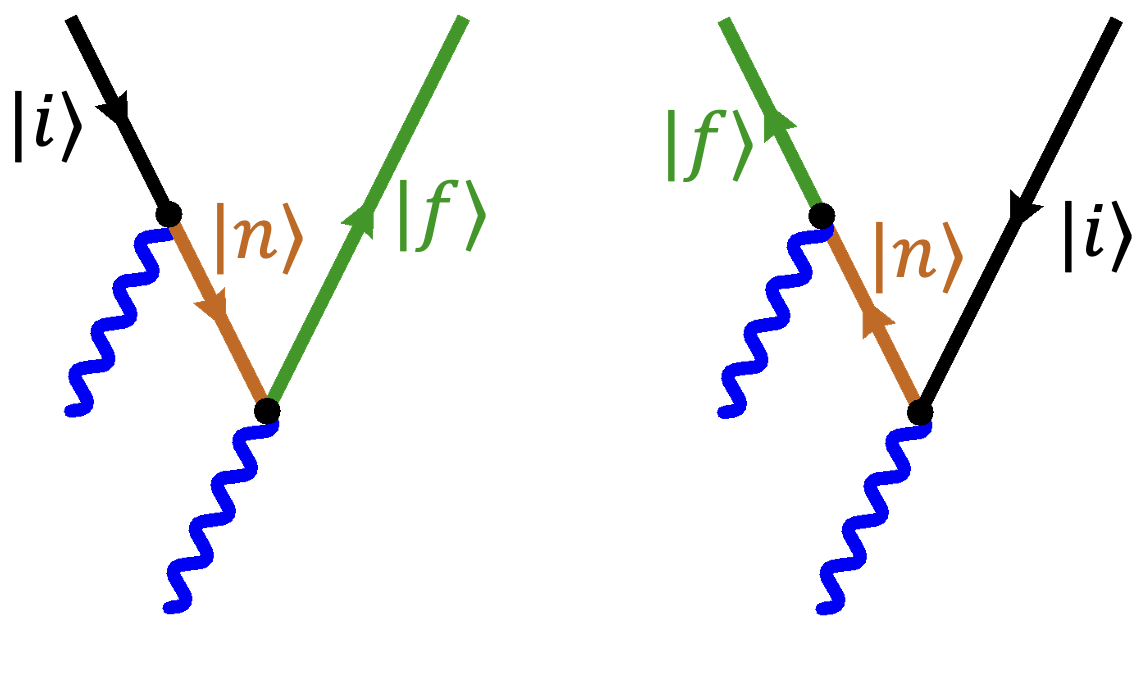}
\caption{Feynman diagrams\cite{sakurai1967} for the TPA amplitude: Incident photons are represented by wavy lines (blue),  the single particle state $|i\rangle$  by  the black line, the the photoelectron $|f\rangle$  by the green line, and intermediate  states $|n\rangle$  by the orange line. The left diagram indicates occupied intermediate states, while the right diagram indicates unoccupied intermediate states.
}
\label{fig:TPA-diagrams}
\end{figure}

Two-photon 
absorption (TPA) and emission (TPE) processes were  originally predicted theoretically by Maria Goeppert-Mayer in her doctoral dissertation.\cite{goppert1931,goppert-translation2009}  
However, TPA 
was not observed until lasers became available and then only for optical frequencies.\cite{Kaiser1961}   
More recently, TPA of hard x-rays  has been observed for metallic
Cu using intense x-ray free electron laser (XFEL) sources.\cite{tamasaku2018}  Formally, the theory of TPA is based on a  sum  of amplitudes for two successive dipole transitions over all possible intermediate states. This sum is given  by the Kramers-Heisenberg (KH) equation. Energy is   conserved only for the net transition, with the transition energy equal to the sum of the two photon
energies.\cite{goppert1931,Mcquire1981,novikov2000,stohr2023}  
This process is  illustrated by the Feynman diagrams\cite{sakurai1967} in Fig.~\ref{fig:TPA-diagrams}. 
While the KH approach is tractable for atomic systems,\cite{Mcquire1981,novikov2000} and non-linear   approaches have been developed for optical spectra,\cite{lam2018}
 quantitative TPA calculations are computationally challenging for condensed matter.   
 However,  for   $K$-shell TPA in Cu with $ \approx$~4500~eV photons,\cite{tamasaku2018}  only non-resonant intermediate states are possible, greatly simplifying the theory.  For this case an approximation for $K$-shell TPA based on the Bethe-Salpeter Equation (BSE) has been proposed.\cite{vinson2022} 

Our goal here is to develop a real-space Green's function (RSGF) approach for deep core TPA in condensed matter that only includes non-resonant contributions and is applicable for simulations of XFEL spectra. 
We show  that this method can be expressed in a form analogous to one-photon (OPA)  x-ray absorption spectra (XAS), but with an effective static quadrupole transition
operator.  
TPA calculations are presented based on an extension of the
RSGF XAS code {\sc FEFF10}.\cite{kas2022}  XFEL effects on the near-edge are modeled by a partially depleted $d$-band.    This theory   yields $K$-shell TPA spectra for Cu in good  agreement with XFEL experiment\cite{tamasaku2018} and with the BSE approach.\cite{novikov2000}

\section{TPA Theory}
 
 Within 2nd-order perturbation theory in the electron-photon interaction, the TPA cross-section $\sigma^{2P}_\mathrm{XAS}$ is given by the many-body Kramers-Heisenberg (KH) equation\cite{novikov2000,stohr2023}
\begin{eqnarray}
\label{eq:mbtpa}
   \sigma^{2P}_\mathrm{XAS}
(\omega\!)\! &=&\!\! 8 \pi^3 \alpha^2 \omega^2
\times \nonumber \\
&&\sum_F |M_{IF}(\omega)|^2  \delta_{\Gamma_F}(2\omega\!+\!E_I\!-\!E_F), \\   
 M_{IF}(\omega)\! &=& \! \sum_X \frac{\langle F|\hat d |X\rangle\langle X|\hat d |I\rangle }{ \omega\!+\!E_I\!-\!E_X\!+\!i\Gamma_X}         .
\end{eqnarray}  
Here 
$|I\rangle$ and $|F\rangle$ denote the initial and final $N$-electron  states  of the system with energies $E_I$ and $E_F$ respectively; $\alpha$ is the fine structure constant; $\hat d=\sum_i d_i$ is the many-body dipole operator in position-space, where  $d_i= {\bf r}\cdot \hat\epsilon$
is the interaction with monochromatic photons of energy $\omega$ and polarization $\hat\epsilon$;  
$|X\rangle$ represent intermediate 
states with energies $E_X$; and the energy denominator
 $\omega+E_I-E_X+i\Gamma_X$
 includes a lifetime broadening $\Gamma_X$. Finally, $\delta_{\Gamma_F}(E)$ denotes a Lorentzian with broadening $\Gamma_F$ associated with the final state lifetime. Unless otherwise specified we use atomic units (a.u.)  
$e=\hbar=m=1$, with energies in Hartrees, distances in Bohr
$a_0=0.529$ \AA, and time in atomic units 24.2 as/Hartree.

The derivation of  TPA theory is similar to other applications of the KH equation,
such as photon-photon scattering,\cite{sakurai1967} with analogous Feynman diagrams. Likewise, contributions from the ${\bf A}^2$ term (i.e., the square of the vector potential) in the electron-photon interaction  known as the {\it seagull} diagram,\cite{sakurai1967} are ignored since they  are arguably of   negligible 
importance.\cite{tamasaku2014x} While typically expressed in terms of dipole operators in position-space ${\bf r}\cdot\hat\epsilon$,  the formulation with momentum-space  ${\bf p}\cdot\hat\epsilon$  operators  is similar, apart from the $\omega^2$ prefactor in Eq.\ (\ref{eq:mbtpa}).\cite{Jortner1967} In general both
resonant intermediate states,  where the energy denominator
vanishes apart from a     
lifetime broadening $\Gamma_X$, and non-resonant intermediate 
states, where the energy denominator is non-zero,
must be considered. When present, resonant states typically dominate, 
and interference between resonant and non-resonant terms
may also be necessary.

For $K$-shell TPA for example, $|I\rangle \approx |1s,N-1\rangle$ represents the initial state prior to x-ray absorption where unless otherwise  specified, the configuration of the  $N-1$ passive electrons is implicit, and in this work also reflects XFEL excitation effects on the system. The final states of interest   
 $|F\rangle=|\overline{1s}\,k L\rangle$  
 have a $1s$ hole (denoted with an overline) and a photoelectron  with wave number $k$, energy $\epsilon_k$, and orbital angular momentum quantum numbers $L=(l,m)$. For simplicity  we  ignore spin indices as well as the  differences between $p_{1/2}$ and $p_{3/2}$ spectra; however, these effects are included implicitly in our numerical calculations. Dipole selection rules  $l\rightarrow l\pm1$  then require intermediate states $|X\rangle$ of two types:
i) states $|\overline{np}kL\rangle$, with a hole in an $np$ orbital and a photoelectron with $d-$ or $s-$symmetry; and ii) states $|\overline{1s}kp\rangle$, with a $1s$ core-hole and a virtual photoelectron with $p-$symmetry. Thus the selection rules for TPA 
are similar to those of normal quadrupole excitations $\Delta l=0,\pm 2$, except that $l_i=l_f=0$ is allowed, and there is an additional restriction $\Delta m=0$.\cite{bonin1984}
In the zero-temperature equilibrium state, near-edge  TPA produces photoelectrons in the $3d$ or $4s$ levels above the Fermi energy $\epsilon_F$.
Neglecting relaxation of the passive electrons (which is discussed  below), the transition energy $E_F-E_I = \epsilon_k-\epsilon_{1s}= 2\omega$. The theory can be extended straightforwardly  to other edges. 

The RSGF approach is based on an independent-electron approach. 
Upon integrating over the $N-1$ passive electrons and rearranging terms,
the KH expression for the TPA cross-section for a given core-level $i$  can be expressed in terms of single-particle levels, which are denoted with lower-case italic indices    
\begin{eqnarray}
  \sigma^{2P}_\mathrm{XAS}
(\omega\!) &=&\! 8 \pi^3 \alpha^2 \omega^2\times \nonumber \\ 
&&\sum^{unocc}_f |M_{if}(\omega)|^2  \delta_{\Gamma_i}(2\omega\!+\!\epsilon_i\!-\!\epsilon_f),  \label{eq:sptpa} \\
M_{if}(\omega) &=&  \sum_{n}
\frac{\langle f| d  |n \rangle \langle n |d | i\rangle}{\omega+\epsilon_{i}-\epsilon_{n}+i\Gamma_n} .\label{eq:Mif}
\end{eqnarray}
Here $i$ and $f$ denote initial-hole and final-photoelectron levels, 
and $\omega=(\epsilon_f-\epsilon_{i})/2$ for  photons of equal energy and $n$ ranges over all intermediate states. This relation for the independent particle transition matrix elements $M_{if}$ requires a careful treatment of the various states involved in TPA along with Fermion commutation relations, and is derived in Appendix A, following the approach of Guo.\cite{Guo1987} 
  In the independent particle formulation, the two types of intermediate levels $n$ correspond to: 1) occupied levels of the ground state allowed by dipole selection rules when the first photon excites an electron in level $n$ to the photoelectron state $f$ and the second excites an electron in the initial level $i$ to level  $n$, or 2) unoccupied levels allowed by dipole selection rules when the first photon excites an electron in the initial core level $i$  to unoccupied level $n$ and the second scatters that electron to the final level $f$.  These processes are illustrated by the two diagrams Fig.\ \ref{fig:TPA-diagrams}. Combining both processes, and assuming that the intermediate state broadening $\Gamma_n\approx\Gamma$ is independent of state $n$, the sum over intermediate levels $n$ in  Eq.\ (\ref{eq:Mif}) is equivalent to the spectral representation of the one-particle Green's function $G(\omega+\epsilon_i)$.\cite{Guo1987} Thus 
 the transition matrix can be represented compactly as
\begin{equation}
M_{if}(\omega) = \langle f| d\,   G(\omega + \epsilon_{i})\, d  | i\rangle,\label{eq:Mif_G}
\end{equation} 
where $G(\epsilon)\equiv [\epsilon-h + i\Gamma]^{-1}$ and $h$ is the independent-electron Hamiltonian for the intermediate states. 

As in OPA, the sum over final states $f$ in Eq.\ (\ref{eq:sptpa}) can be treated implicitly using  Green's function methods, \cite{kas2022,ankudinov1998}
 in terms of the  one-particle density matrix 
\begin{equation}
   \hat\rho(\epsilon) \equiv -\frac{1}{\pi}{\rm Im}\, G(\epsilon) 
   \equiv\sum_f |f\rangle \delta(\epsilon-\epsilon_f)\langle f|.
   \label{eq:spectraldensity}
\end{equation}
Here $f$ are the  eigenstates of the final-state independent particle Hamiltonian $h'$ in the presence of the screened core hole.  In order to restrict the sum in Eq.\ (\ref{eq:sptpa}) to unoccupied levels, a factor ${\overline g}(\epsilon)=1-g(\epsilon)$ is appended, that characterizes the distribution.  
 In the ground state ${\overline g}(\epsilon)= \theta(\epsilon_F-\epsilon)$ is just a step function at the Fermi energy.  
 However, in general ${\overline g}(\epsilon)$ is a non-equilibrium distribution  \cite{mercadier2024} that  depends on details of the experiment and the  x-ray source. For example, in transient XAS studies of warm dense matter (WDM), at fs time-scales  the behavior of  $g(\epsilon)$ has been addressed using Boltzmann equation methods.\cite{mercadier2024} At longer ps time-scales,  $g(\epsilon)$ is taken to be the Fermi function $f(\epsilon,T)=1/[\exp((\epsilon-\mu)/k_BT) +1]$ at the transient electron temperature
$T$ in the system, where $k_B$ is the Boltzmann constant.\cite{cho2011electronic,tan2021}  Due to the dependence on depletion, the TPA cross section below the cold edge will not simply be proportional to intensity squared, since the depletion of the occupied orbitals is already proportional to intensity, giving a cubic dependence. The current experimental results were extracted via quadratic fitting analysis, and thus the depletion factor used in our theory can be viewed as an average over the intensities probed in the experiment.
 
 From Eq.\ (\ref{eq:sptpa})-(\ref{eq:spectraldensity}) the TPA can be expressed   as 
\begin{eqnarray}  
\sigma^{2P}_\mathrm{XAS}(\omega)&=& 8 \pi^3\alpha^2\omega^2\langle i| \tilde Q^\dagger(\omega) \,\tilde\rho(\epsilon)\, \tilde  Q(\omega) |i\rangle, \label{eq:2pxas}\\
 \tilde Q(\omega) &=&  d\, G(\omega+\epsilon_{i})\, d ,\label{eq:qomega} \\
 \tilde\rho(\epsilon)&=&\int d\epsilon'\  A_{\Gamma_i }(\epsilon - \epsilon')\hat\rho(\epsilon')\bar g(\epsilon'),\label{eq:rhotilde}
\end{eqnarray}
where $\epsilon = 2\omega+\epsilon_{i}$, $\tilde Q(\omega)$ is 
a dynamic quadrupole operator, and the convolution over the Lorentzian $A_{\Gamma_f(\epsilon)}$ produces the final state lifetime broadening. This representation of the TPA   is valid for  both   resonant and non-resonant TPA. Thus Green's function methods have been used to calculate TPA both in model and atomic systems.\cite{mcguire1981} 

\section{Non-resonant TPA }

 Our primary focus in this work is non-resonant $K$-shell TPA in Cu with monochromatic linearly polarized photons of energy $\omega$ near 4485 eV,
 and polarization $\hat\epsilon = \hat z$, 
 as in the   XFEL  measurements.\cite{tamasaku2018} To simplify the analysis, we choose polarization $\hat\epsilon=\hat z$ 
 without loss of generality, as there is no polarization dependence in isotropic systems like Cu. In this case the TPA involves a 2-photon transition from a deep $1s$ level and energy $\epsilon_{1s}\approx -8970$~eV to 
 a  photo-electron in level  $|f\rangle = |kL\rangle$ with   angular momentum indices $L=(l,m)$ for $l=0$ or 2, and energy $\epsilon_k$ in the unoccupied levels. Due to transient effects of   short high intensity XFEL pulses ${\overline g}(\epsilon)$ is a non-equilibrium distribution that characterizes a partially depleted $d$-band in Cu.\cite{mercadier2024}  

  For $\approx$ 4500 eV photons no resonant intermediate states exist for which the energy denominator  $D_n(\omega)=\omega+\epsilon_{1s}-\epsilon_n=0$ in the core-spectrum between the 1s level at $-$8960 eV and $\epsilon_F$.
Dipole selection rules imply that the only available intermediate levels  $n$  are of $p$-symmetry with  $n\geq 2$, both below and above $\epsilon_F$.
At the onset of TPA  $2\omega \approx$ 8970 eV, and  $\omega+\epsilon_{1s}\approx -\omega$. Thus the matrix elements in non-resonant $K$-edge TPA  in Cu depend  on the behavior of the diagonal Green's function elements with angular momentum $l=1$ at large negative energies $G_1(-\omega)=G_{l=1}(-\omega)$.   
From its spectral representation, contributions to $G_1(\epsilon)$ from the bound $np$ levels below    $\epsilon_F$ denoted by $G_1^<(\epsilon)$, can be expressed as a sum over the  $2p$ and $3p$ levels. The sum over the $p$-level continuum  above $\epsilon_F$ denoted by $G_1^>(\epsilon)$, is less straightforward. 
However, as argued by Vinson,\cite{vinson2022} the energy denominator
$D_{np}(\omega) =\omega+\epsilon_{1s}-\epsilon_{np}=\epsilon_f-\epsilon_{np}-\omega$ is nearly constant over several hundred eV for the continuum
$np$ levels above $\epsilon_F$,  and  can be approximated by $D_0=-\omega$. 
Then using the projection operator ${\bf P}_p$  onto occupied $p$-levels,  the sum can be approximated by
\begin{equation}
G_1^>(-\omega) = \sum_{np>3} \frac{ |np\rangle\langle np| }{D_{np}(\omega)}  \approx
\frac{ {\bf 1}-{\bf P}_p}{D_0},\label{eq:d0}
\end{equation}
where $\bf{1}$ is the unit operator. The error in this approximation is of order $(\epsilon_k-\epsilon_F)/\omega^2$ near $\epsilon_F$, where $\epsilon_k$ is the photoelectron energy, and hence is negligible for photoelectron energies $\epsilon_k$ near the TPA $K$-edge  in Cu. The approximation in Eq.\ (\ref{eq:d0}) can be verified by replacing the unit operator with a complete set of one-particle levels and appropriate selection rules. 
As a consequence, the the $np$ states above  $\epsilon_F$ tend to cancel those from the occupied 2$p$ and 3$p$ levels. Combining these terms, $G_1(-\omega)$ for Cu can be approximated   by
\begin{equation}
    G_1(-\omega)\approx -\frac{1}{\omega}\left[ 1 + \sum_{np}^{n=2,3} C_{np} |np\rangle\langle np| \right],\label{eq:G1approx}
\end{equation}
where $C_{np} = D_0/D_{np} - 1$. For 4485 eV photons
$C_{2p} \approx - 4485/(-4485+960)-1 \approx 0.27$, and 
$C_{3p} \approx -4485/(-4485 +70)-1 \approx 0.016$.   

From Eq.\ (\ref{eq:2pxas}), (\ref{eq:qomega}) and (\ref{eq:G1approx}), the  TPA  can be expressed in terms of a
static quadrupole  transition operator $Q$  as
\begin{eqnarray}
 \sigma^{2P}_\mathrm{XAS}(\omega) &=& 8 \pi^3\alpha^2 \, \langle 1s|   Q  \,
 \tilde \rho(\epsilon)
 \, Q  |1s\rangle, \\
Q &\equiv& 
    d^2   + \sum_{np}^{n=2,3} C_{np}\,
    d  |np\rangle\langle np|\,d  , 
\label{eq:Q_static}
  \end{eqnarray} 
where $\epsilon = 2\omega+\epsilon_{1s}$.
Note that the
factors $ \omega^2$ from the matrix elements defined in  Eq.\ (\ref{eq:G1approx})   cancel the $\omega^2$ prefactor in Eq.\ (\ref{eq:sptpa}). This yields, 
\begin{align}    
   \sigma^{2P}_{XAS}(\omega) &=   {8\pi^3 \alpha^2 }  \sum_L^{l=0,2}
   { |M^Q_L(\epsilon)|^2}  \tilde \rho_l(\epsilon). 
\label{eq:tpacompact}
\end{align}
Here   
 $M_L^Q(\epsilon) = \langle 1s| Q | k L\rangle$
 is the quadrupole transition matrix, and $\tilde \rho_l(\epsilon)$ is the $l$ component of the density matrix defined in Eq.\ (\ref{eq:rhotilde}). 
The expression in Eq.\ (\ref{eq:tpacompact})  for non-resonant TPA $\sigma^{2P}_\mathrm{XAS}(\omega)$   is a key result of this paper. 
Analogous effective quadrupole operators have been used   in related contexts, e.g.,  double $\gamma$ decay in nuclei.\cite{Bertsch1972}

The treatment of the near-edge TPA with XFEL radiation depends on several factors.  At low to intermediate intensities, XFEL effects lead to a partial depletion of the $d$-band and the emergence of spectra below the cold edge.\cite{mercadier2024} Although more sophisticated approaches require a Boltzmann equation treatment, for
  simplicity here   we approximate $\overline g(\epsilon)$ by a   step function with $\overline g=0.075$ for $d$-levels below the cold Fermi level and $\overline g\approx  1.0$ above, reflecting a broad distribution of the excited 3{\it d} electrons  across high-energy unoccupied states. We also accounted for experimental broadening by adding 1.2 eV to the core-hole lifetime, which is equivalent to Lorentzian broadening.  Core-hole screening is also an important  consideration. For $K$-shell TPA in Cu, we find that RSGF   
calculations that neglect the core-hole potential in the final state are a good approximation, similar to the   behavior of $L_{2,3}$ XAS in metals.

For comparison with the RSGF approach outlined in Eqs.~(3) -- (15),  it is useful to compare with the BSE.
The key difference is that the BSE uses an electron-hole basis denoted by the composite index  $b=\{e,h\}$,
where the photo-electron inhabits unoccupied orbitals $e$ and the hole occupied core levels $h$.  
However,   the BSE also uses  the  same approximation for the effective quadrupole operator $Q$ in Eq.\ (\ref{eq:Q_static}). The BSE analog of Eq.\ (\ref{eq:tpacompact}) is
\begin{align}
&\sigma^{2P}(\omega) =    \frac{ 8 \pi^3 \alpha^2}{\Omega N_k} \times  \label{eq:tpa_bse} \\
& \times \left(-\frac{1}{\pi} \mathrm{Im}\right)  \left[ \sum_{bb'} \langle i \vert Q \vert b \rangle  \langle b \vert \frac{1}{2\omega - H_{\mathrm{BSE}} + i\Gamma} \vert b' \rangle \langle b' \vert Q \vert i \rangle \right]. \nonumber 
\end{align}
Here the integral over the Brillouin zone is replaced by a finite sum over $N_k$ points, $\Omega$ is the unit cell volume,  
and the effective  BSE particle-hole  Hamiltonian is   
\begin{equation}
H_{\mathrm{BSE}} = \epsilon_e - \epsilon_h + \sqrt{1-g(\epsilon_{e})} \left( V_X - W \right) \sqrt{1-g(\epsilon_{e'}}) \;,
\label{eq:bse}
\end{equation}
where $\epsilon_e$ and  $\epsilon_h$ are the energies of the non-interacting electron and hole, $V_X$ is the exchange interaction, $W$ is the direct interaction, and $g(\epsilon)$ is the occupancy of the transient, non-interacting electron states, as discussed in Ref.~\onlinecite{vinson2022} and  \onlinecite{PhysRevB.93.195205}.  
Similar to the RSGF approach, the occupation of the upper valence bands (Cu 3{\it d} and 4{\it s}) was reduced to $g(\epsilon)$=0.925 to account for $d$-band depletion caused by the XFEL. 
In Eq.~(\ref{eq:bse}) the effect of the core-hole potential and exchange are suppressed as $g(\epsilon) \rightarrow 1$. 

\section{Calculations}

Calculations of the TPA using Eq.\ (\ref{eq:tpacompact}) are carried out using a straightforward  extension of the RSGF approach for XAS in {\sc FEFF10},\cite{kas2022,ankudinov1998} by substituting the dipole  operator $d$ with $Q$.
As in XAS, the one-particle Green's function  
for these calculations is based on a quasi-particle approximation that builds in a self-energy and many-body final state effects. That is,
    $h=h'+\Sigma(\epsilon)$, where $h'$ is the final-state Hartree Hamiltonian in the presence of a  screened core-hole,  and  $\Sigma(\epsilon)$ is the electron self-energy (i.e., the dynamic exchange-correlation potential), which implicitly includes
  an additional imaginary part representing the 1$s$ core-hole lifetime. The Hamiltonian $h'$ implicitly assumes a given electron configuration, in this case that with the partially depleted $d$-band. 
  The RSGF calculations are then carried out using
  spherical muffin-tin potentials.\cite{ankudinov1998} The density matrix at the absorbing atom is represented in terms of the local site-angular momentum scattering-states $R_L({\bf r},\epsilon)$  
\begin{eqnarray} 
    \rho({\bf r}, {\bf r}',\epsilon) &=& 
-\frac{1}{\pi}{\rm Im}\, G({\bf r},{\bf r}',\epsilon), \\
  &=&  \sum_L R_L ({\bf r},\epsilon) R_L^*({\bf r}',\epsilon)\rho_l(\epsilon),
\end{eqnarray}
where $\rho_l(\epsilon)$ is the local projected density of states for angular momentum $l$. For convenience in FEFF10, a factor $2k/\pi$ that accounts for spin degeneracy and the density of continuum levels is lumped into the normalization of radial wave-functions.\cite{ankudinov1998}  With this  convention   $\rho_l(\epsilon) = 1 + \chi_l(\epsilon)$,
where $\chi_l(\epsilon)$ is the fine-structure for a given angular momentum  component  $l$
arising from multiple-scattering from neighboring atoms.

   The matrix elements $M_L^Q(\epsilon)$ are  calculated  by first expanding $Q$, $R_L({\bf r})$, and $R_{1s}({\bf r})$ in spherical harmonics,  
   and integrating over all angles. For a $1s$ initial state, $L$ is that for the photoelectron level $|kL\rangle$. 
   The quadrupole matrix elements $M^Q_L(\epsilon)$ are then given by radial integrals
   \begin{align}
       M^Q_L(\epsilon) =q_l\langle R_{l}(r)| \big[ &r^2 R_{1s}(r) + r C_{2p}m_{1s}^{2p}  R_{2p}(r)  \nonumber \\
       &+ r C_{3p}m_{1s}^{3p}   R_{3p}(r) \big] \rangle,\label{eq:radial_me}
   \end{align}
   where $q_l=\langle Y_L^* Y_{10}Y_{10} Y_{00} \rangle$, and $m_{1s}^{np}=\langle R_{np}| r |R_{1s}\rangle$ are radial dipole matrix elements.
    The terms in Eq.\ (\ref{eq:radial_me}) are then calculated using  an extension of the the non-resonant inelastic scattering (NRIXS) module 
in the {\sc FEFF10}  code.\cite{soininen2005}
Inserting these terms into Eq.\ (\ref{eq:tpacompact}) then yields all contributions to $|M_L^Q(\epsilon)|^2$ including cross terms.  
 Finally we considered different models for the screened core-hole in $K$-edge TPA, and found that, similar to  $L_{23}$ OPA in metals, 
 to a good approximation the core-hole can simply be ignored. This is done with a {\it NOHOLE} setting in {\sc FEFF10}.  For comparison we also show results from the BSE approach of Eq.\ (\ref{eq:tpa_bse}), where  to account for $d$-band depletion caused by the XFEL, the occupation of the upper valence bands (Cu 3{\it d} and 4{\it s}) was reduced to $g(\epsilon)=0.925$ along with 1.2 eV broadening. 
 
\begin{figure}[t]
\includegraphics[trim={50 50 70 30 },width=1.0\columnwidth]{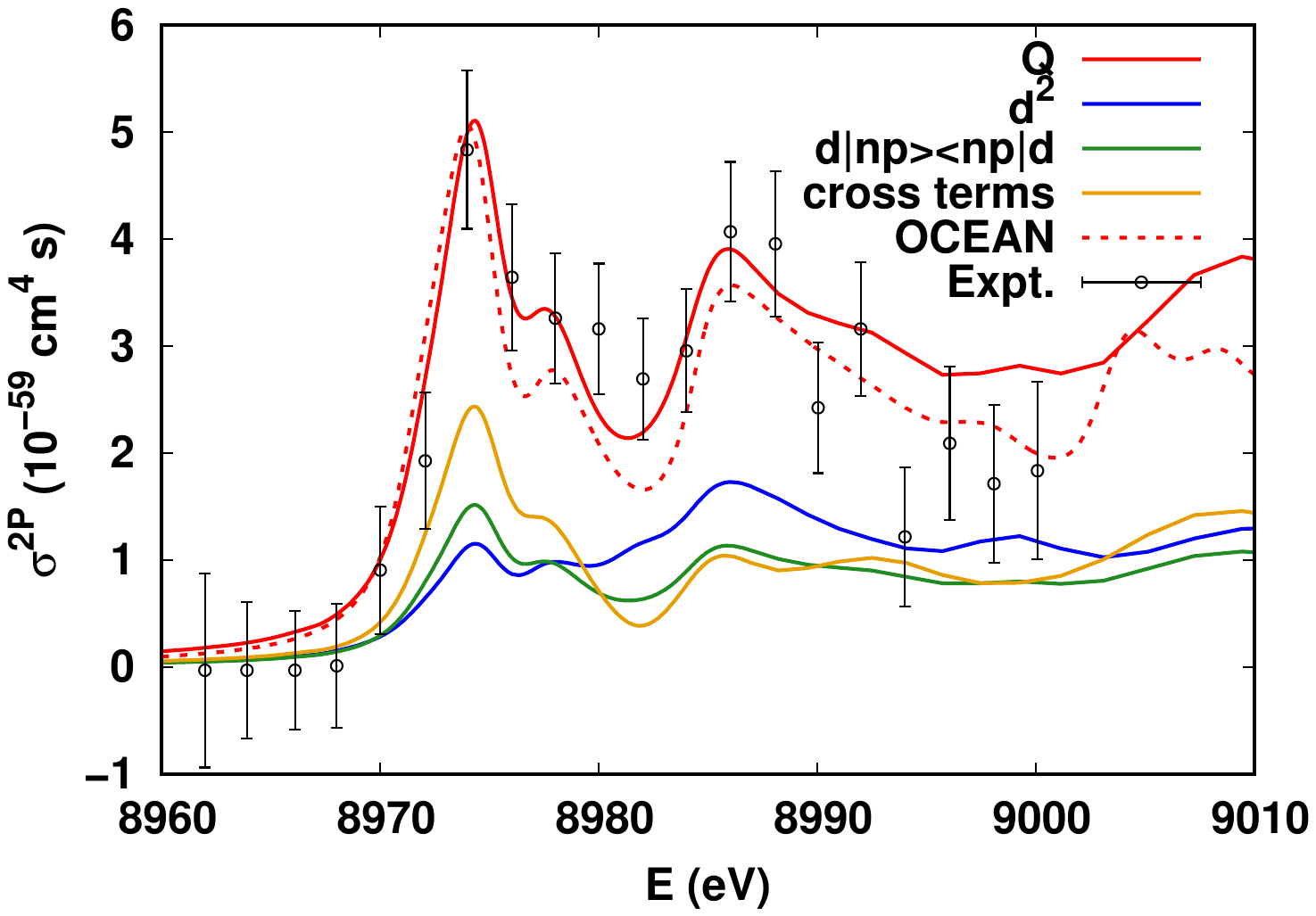}
\caption{$K$-edge two photon absorption spectrum of Cu calculated using RSGF and Eq.\ (\ref{eq:tpacompact}): full calculation with quadrupole operator $Q$ (red), pure quadrupole operator $d^2$ (blue), $2p$ and $3p$ contributions
(green), cross-terms (gold), and for comparison, results from Eq.\ (\ref{eq:tpa_bse}) using the {\sc ocean} BSE code (red-dashed), and from experiment\cite{tamasaku2018} (black). 
\label{fig:TPA-Cu-FEFF10}
} 
\end{figure}

 Our RSGF  results are presented  in Fig.\ \ref{fig:TPA-Cu-FEFF10}.
Clearly, the full TPA calculation  is in quantitative agreement with experiment, within error bars. The RSGF results are also  nearly the same as those    with the BSE. This agreement indicates that our approximation of the $d$-band depletion by a   step function is reasonable.  
Interestingly, all of the   contributions to TPA in   $|M_L^Q|^2$ including cross-terms are of   comparable magnitude. Although the coefficients $C_{2p}$ and $C_{3p}$ are small, these factors are compensated by the matrix elements $m_{1s}^{np}$ that depend  on the longer range  $2p$ and $3p$  wave-functions. 

\begin{figure}[ht]
\includegraphics[trim={50 50 70 30 },width=1.0\columnwidth]{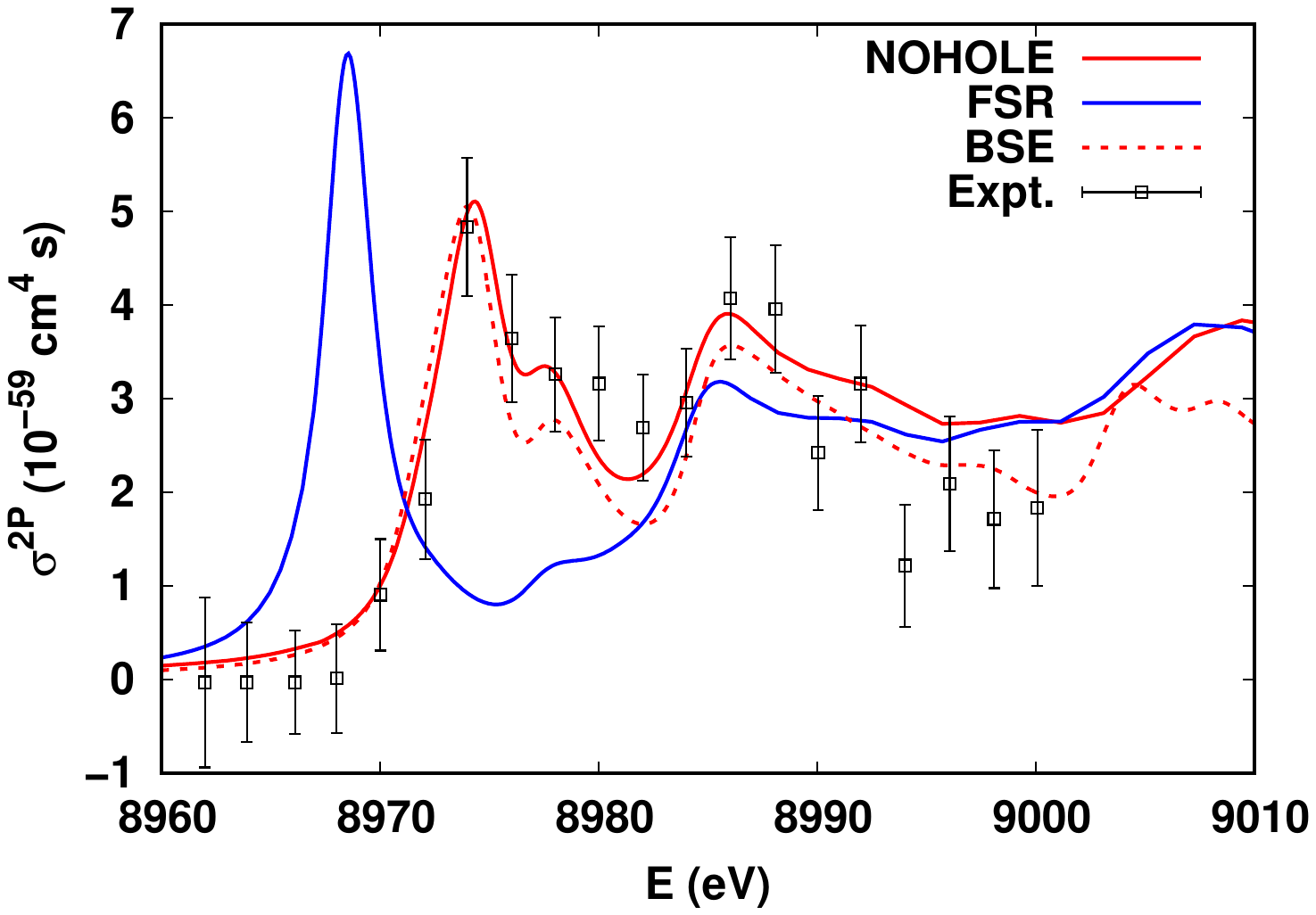}
\caption{$K$-edge two photon absorption spectrum of Cu calculated using RSGF and Eq.\ (\ref{eq:tpacompact}): full calculation with either no core-hole (red) or a DFT screened full core-hole (FSR)     (blue), compared to experiment\cite{tamasaku2018} (black). and   the  BSE code {\sc ocean}(red-dashed).  Note that  FSR screening is too strong and leads to a large unphysical peak below the edge. } 
\label{fig:TPA-Cu-FEFF10-FSR}
\end{figure}

As a check on the effect of the core-hole, we   carried out calculations with a density functional theory (DFT) screened core-hole as in the final state rule (FSR), as shown in Fig.\ \ref{fig:TPA-Cu-FEFF10-FSR}. 
This approximation sharpens the pre-edge peak, and produces a red shift of the peak by $\approx 5$~eV, in significant disagreement with experiment.
\begin{figure}[ht]
\includegraphics[trim={50 50 70 30 },width=1.0\columnwidth]{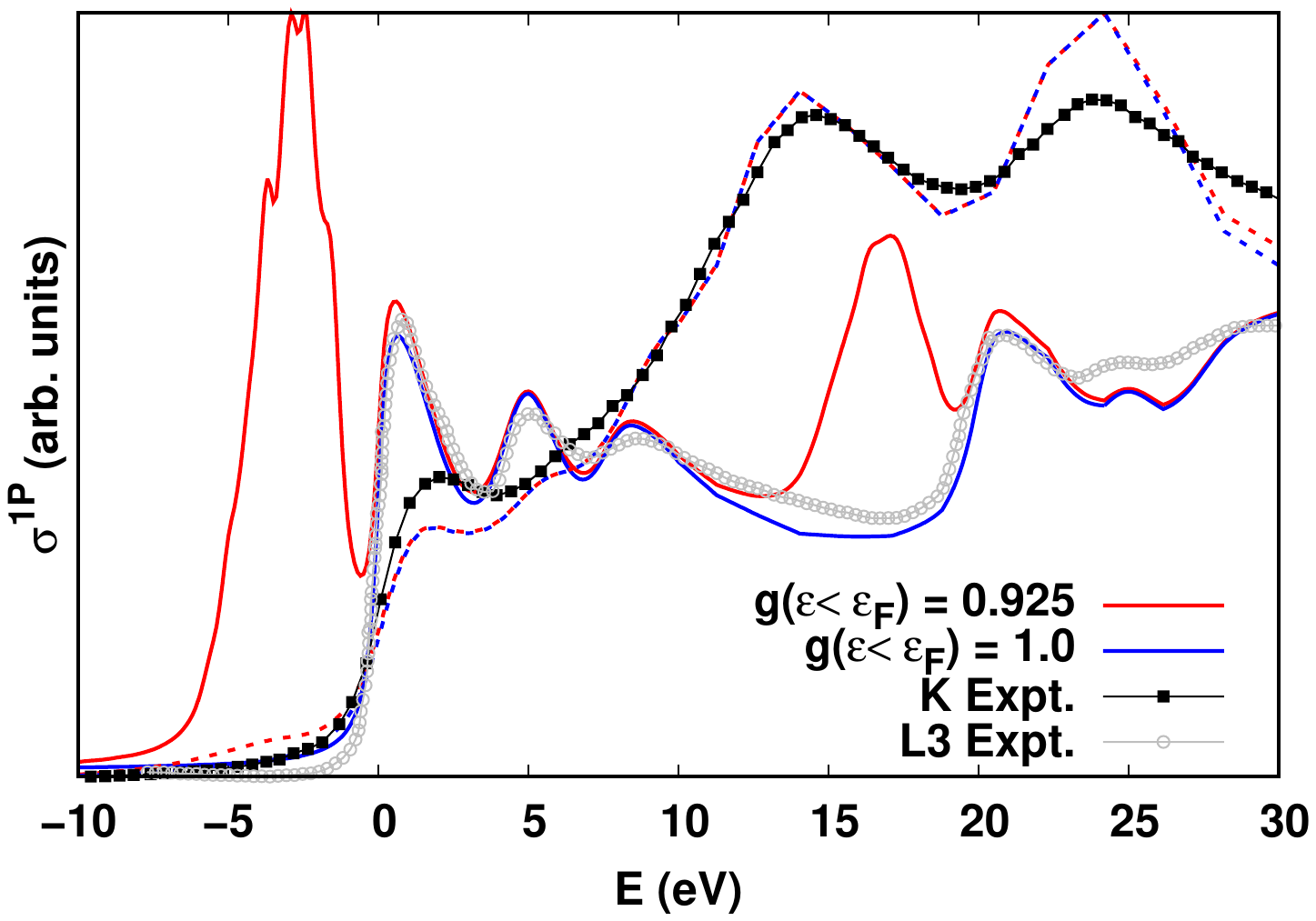}
\caption{Cu K-edge (dashed) and L$_{23}$-edge (solid) OPA
 XANES as a function of energy relative to the cold edge $E=0$, calculated using RSGF with partial occupation (red) and full occupation (blue) of the upper valence states compared to room temperature experimental data (dots and crosses).\cite{ebert1996}} 
 \label{fig:TPA-Cu-FEFF10-OPA}
\end{figure}
  We also checked that RSGF calculations of $K$-edge OPA for Cu using Eq.\ (\ref{eq:Q_static})  and  similar occupation factors $g(\epsilon)$ have very small contributions from  analogous $p$-band depletion below the cold Fermi level from XFEL sources (Fig.\ \ref{fig:TPA-Cu-FEFF10-OPA}). This difference is due to the much smaller $p$-partial density of states in Cu,  consistent with  experimental $K$-edge XFEL OPA for Cu.\cite{tamasaku2018}

\section{Interpretation and Discussion}

The expressions for the TPA in Eqs.\ (11) and  (15) are  
formally similar to the golden rule for OPA $\sigma^{1P}_\mathrm{XAS}(\omega)$, except for the replacement of the dipole operator $d$ by the quadrupole operator $Q$  
\begin{eqnarray}
   \!\! \sigma^{1P}_\mathrm{XAS}(\omega)\!&=&\! 4 \pi^2\alpha \omega
    \sum^{unocc}_{kL} |\langle kL| d |1s\rangle|^2 \delta_{\Gamma_{1s}}(\omega \!+\epsilon_{1s}\! -\!\epsilon_k) \quad \\  
   \!\! \!&=&\! 4 \pi^2\alpha \omega\, \langle 1s| d\, \tilde\rho(\epsilon)\, d |1s\rangle, 
\end{eqnarray}
where $\epsilon =\omega+\epsilon_{1s}$ and $\tilde \rho(\epsilon)$ is given by Eq.\ (\ref{eq:rhotilde}).  
 
 Consequently the $K$-shell TPA in Eq.~(\ref{eq:tpacompact})
   is  formally analogous to  $L_{23}$ OPA, i.e.,
\begin{equation}
\sigma_{L_{23}}^{1P} (\omega) = 4\pi^2\alpha \omega \sum_L^{l=0,2}
|M_L(\epsilon)|^2\,  \tilde\rho_l(\epsilon)  
\label{eq:opaL3}
\end{equation}
where  $\epsilon=\omega + \epsilon_{2p}$.  The main differences are the quadrupole transition operator $Q$ in $|M_L^Q(\epsilon)|^2$ which leads to quadrupole selection rules. The structure of $Q$ has terms from intermediate levels both above and below $\epsilon_F$, corresponding to the  contributions from the occupied and unoccupied intermediate states in the Feynman diagrams in Fig.\ \ref{fig:TPA-diagrams}.   Also, both $K$-edge TPA and $L_{23}$ XAS have 
the same $s$- and $d$-densities of continuum states $\rho_l(\epsilon)$. However the atomic background from the $2p$ contribution to $|M_L(\epsilon)|^2$ 
is not identical and the edge onset is modified in the XFEL experiment due to the $d$-band depletion, in contrast to ground state calculations. 

The expression for $\sigma^{2P}_K(\omega)$  from  pure quadrupole contribution $d^2$ to  $Q$,  is   related to the   $K$-edge  quadrupolar OPA cross-section $\sigma_{K}^{1P-Q} (\omega_K)$ for photon wave vector $\bm{\kappa}=(\omega_K/c)   \bm{   \hat \epsilon}$ at energy $\omega_K=2\omega$,
\begin{equation}
\sigma_{K}^{1P-Q} (\omega_K) = 4\pi^2\alpha\omega_K \sum_L^{l=0,2}
|M^{Q_{1P}}_L(\epsilon)|^2\tilde \rho_l(\epsilon) 
\label{eq:OPAquad}
\end{equation}
where $Q_{1P}=\sfrac{i}{2} (\bm{\kappa\cdot r})(\bm{\epsilon\cdot r})$, and for $K$-edge XAS $\omega_K=\epsilon-\epsilon_{1s}=2\omega$. Comparing Eq.\ (\ref{eq:tpacompact}) and (\ref{eq:OPAquad}) yields
\begin{equation}
    \sigma_{K}^{2P} (\omega) \approx \frac{2\pi\alpha}{\omega_K\kappa^2} \sigma_{K}^{1P-Q} (\omega_K) 
\label{eq:Ktpaopa}
\end{equation}
However, there is no contribution to $l=0$ $s$-states from $Q_{1P}$, in contrast to $K$-shell TPA. In addition, the added weight from the occupied intermediate states in TPA is appreciable. 

Though relatively small for Cu,   relaxation of the passive $N-1$ electrons can   be important in TPA.\cite{novikov2000} These relaxation effects 
account   for  shake-processes, edge singularity effects, and energy shifts due to the sudden turn-on of the 1$s$ core-hole.  In principle such effects can  be added   using approximations similar to those for XAS.\cite{campbell2002} In this approach, the $G(\omega)$ in 
Eq.~(\ref{eq:spectraldensity}) is replaced by an effective Green's function $G_{\rm eff}(\omega)\equiv A*G_{qp}$ that includes a convolution  with the core-hole spectral function $A_c(\omega,\omega')$, similar to that for XAS. This spectral function can be approximated, e.g., using cumulant Green's function methods.\cite{campbell2002,kas2022} Moreover, the transition matrix elements must be modified to include a projector onto unoccupied one-particle states of the ground state Hamiltonian $h$, i.e.,
${\overline M}^Q_{if} = \langle 1s | Q {\overline P} |k'L\rangle$, where $\epsilon_k'$ are the single-particle energies in the presence of the screened $1s$ core hole. This projector accounts for the Mahan   enhancement factor at the edge. Since the effects are not expected to be significant for Cu, the implementation of this extension  is reserved for the future.

  From Eq.\ (\ref{eq:tpacompact}), it is straightforward to obtain an approximate sum-rule for the   TPA from a given core-level. $\int d\omega\, \hat \rho(2\omega+\epsilon_{1s}){\overline g}(\epsilon) = 
(1/2) [{\overline{\bf P}}(T)] = (1/2) [{\bf 1}- {\bf P}(T)]$, where ${\bf 1}$ is the unit operator and ${\bf P}(T)$ [${\overline{\bf P}}(T)$] is the projector onto occupied [unoccupied] levels, weighted by $g(\epsilon)$. 
\begin{eqnarray}
      && \int  \!\! d\omega \,  \sigma^{2P}_{1s-XAS}(\omega)   \approx   4 \pi^3
    \alpha^2 \langle 1s |  Q\,{\overline{\bf P}}(T)\,Q^\dagger |1s \rangle \\
 \!\!\!\!   &=&  4 \pi^3
    \alpha^2  \Big[ \langle 1s|QQ^\dagger|1s\rangle \! -\!\!\sum^{l=s,d}_{nl} |\langle 1s| Q |nl\rangle|^2 g(\epsilon_{nl})   \Big]  . \quad\;  
\end{eqnarray}
This result reflects the expectation value of $QQ^\dagger$ in a given core-level minus a correction from the projection of $Q$ onto occupied $s$ and $d$-levels.   
In contrast, the Thomas-Reichi-Kuhn  OPA   sum-rule\cite{sakurai1967}   sums dipole allowed transitions over all occupied levels,
\begin{equation}
    \int d\omega\, \sigma^{1P}_\mathrm{XAS}(\omega) = 2\pi^2\alpha  
      \sum_n^{occ} \langle n| [[h,d],d] |n\rangle =  \frac{2 \pi^2 \alpha \hbar}{m} Z.
\end{equation} 
 The prefactor $ 2 \pi^2 \alpha \hbar/m$ is independent of the system and has a value  $\pi h c r_0 = 0.144\, \mathrm{a.u.} = $ 110 Mb eV, where
 $r_0=e^2/mc^2$ is the classical electron radius.  

\section {Summary and Conclusions}
We have presented a tractable approximation for non-resonant  TPA  based on  an  independent particle Green's function treatment of the non-resonant intermediate states, and we have applied it to the K edge of metallic copper.  Our  formulation  starts with the many-body KH formula  \cite{novikov2000} 
and approximates the sum over
intermediate states  
above the Fermi level using a projection unto occupied levels.\cite{vinson2022} These closure techniques are similar to those of St\"ohr, which assume a constant energy denominator.\cite{stohr2023}    
These approximations then lead to an expression for non-resonant TPA in Eq.\ (\ref{eq:tpacompact}) in terms of an effective,
static quadrupole operator $Q$ and a non-equilibrium occupancy function $g(\epsilon)$ due to XFEL induced depletion of the valence bands.   For $K$-edge TPA in Cu,
$Q$ has only three   terms in Eq.\ (\ref{eq:radial_me}), a pure quadrupole interaction $d^2$ and terms
from the projectors onto occupied $2p$ and 3$p$ states. Our approximation of non-resonant TPA in Eq.~(\ref{eq:tpacompact}) is  implemented using a straightforward extension of the  NRIXS module 
in the RSGF {\sc FEFF10} XAS code. XFEL effects lead to a partially depleted $d$-band which for simplicity is modeled by a   step   function distribution $g(\epsilon)=0.925$ below the cold Fermi level. In addition, we have included 1.2 eV experimental broadening, and a {\it NOHOLE} approximation. While improved  treatments of the distribution function require Boltzmann equation techniques,\cite{mercadier2024} the present model is already in quantitative agreement with experiment within error bars and in absolute units.\cite{tamasaku2018}  
We also find that all contributions including the cross-terms in $|M^Q_L(\epsilon)|^2$ are of comparable magnitude. 

Finally an extension of this work for the  treatment of two-photon x-ray emission (TPE) is  similar in many respects  to that for TPA. The difference is analogous to that between XAS and XES,\cite{ankudinov1998} i.e., the main change is  the replacement of the complementary   distribution function ${\overline g}(\epsilon)$ in Eq.~(\ref{eq:rhotilde}) by $g(\epsilon)$ that  restricts the initial states to occupied levels.   \\

\noindent {Acknowledgments} ---
We thank  G.\ Bertsch for illuminating discussions, R.\ Albers and B.\ Ziaja for comments, and Alex Kennedy for assistance with some of the calculations.
This work is supported in part by the Theory Institute for Materials and Energy Spectroscopy (TIMES) at SLAC, FWP100291 which is funded by DOE Office of Science BSE DMSE Contract DE-AC02-76SF0051.
Certain software is identified for informational purposes. Such identification does imply recommendation or endorsement by the National Institute of Standards and Technology,
\bibliography{references}

\appendix
\section{TPA Transition Matrix Elements}
The single particle matrix elements are derived by  assuming the many-body states are all single Slater determinants. This ignores relaxation of the passive $N-1$ orbitals. 
Following Guo,\cite{Guo1987} we define the intermediate and final many-body states as particle-hole states with $|F\rangle=a^{\dagger}_f a_i |I\rangle$. There are two types of intermediate states in TPA: 1) those with a hole in a core-state $i$ and a particle in an arbitrary unoccupied state $j$, and 2) those with a hole in an arbitrary occupied state $a$ and a particle in state $f$. These intermediate states are given by $|X\rangle=a^{\dagger}_j a_i|I\rangle$ and $|X\rangle=a^{\dagger}_f a_{a}|I\rangle$. Below we  keep this notation with $a$ denoting occupied states, $j$ denoting unoccupied states, and $q,r$ denoting either. In addition, the many-body dipole operator is $D=\sum_{qr}d_{qr}a^{\dagger}_q a_r$, where $d_{qr}=\langle q|d|r\rangle$. Noting that the sum in the dipole operator is always collapsed such that the dipole operator connects the ground state to the intermediate state and the intermediate state to the final state,
one can rewrite the matrix elements as,
\begin{align}
    M_{IF} &= 
    \sum_{j}^{unocc}\frac{\langle I|a^{\dagger}_{i}a_f d_{fj}a^{\dagger}_f a_j a^{\dagger}_j a_{i}|I\rangle\langle I|a^{\dagger}_{i}a_{j}d_{ji}a^{\dagger}_{j}a_{i}|I\rangle}{\omega + \epsilon_i - \epsilon_j} \nonumber \\
    &+ \sum_{a}^{occ}\frac{\langle I|a^{\dagger}_{i}a_f d_{a i}a^{\dagger}_{a} a_i a^{\dagger}_f a_{a}|I\rangle\langle I|a^{\dagger}_{a}a_{f}d_{f a}a^{\dagger}_{f}a_{a}|I\rangle}{\omega + \epsilon_{a} - \epsilon_f} \nonumber \\
    &=\sum_{j}^{unocc}\frac{d_{fj}d_{ji}\langle I| n_{i}\bar n_f \bar n_j|I\rangle\langle I|n_{i}\bar n_{j}|I\rangle}{\omega + \epsilon_i - \epsilon_j} \nonumber \\
    &- \sum_{a}^{occ}\frac{d_{a i}d_{f a}\langle I| n_{i}\bar n_f \bar n_{a}|I\rangle\langle I|n_{a}\bar n_{f}|I\rangle}{\omega + \epsilon_{a} - \epsilon_f} \nonumber \\
    &= \sum_{n}^{unocc} \frac{d_{f j} d_{j i}}{\omega + \epsilon_{i} - \epsilon_j}-\sum_{a}^{occ} \frac{d_{fa} d_{a i}}{\omega + \epsilon_{a} - \epsilon_f},
\end{align}
where $n, \bar n$ are particle- and hole-number operators   respectively. 
Then using conservation of energy $2\omega = \epsilon_f - \epsilon_i$ gives,
\begin{align}
    M_{IF}&=\sum_{j}^{unocc} \frac{d_{f j} d_{j i}}{\omega + \epsilon_{i} - \epsilon_n}+\sum_{a}^{occ} \frac{d_{f a} d_{a i}}{\omega + \epsilon_i - \epsilon_{a}} \nonumber \\
    &= \sum_n \frac{d_{f n} d_{n i}}{\omega + \epsilon_{i} - \epsilon_n},
\end{align}
where the final sum is over all levels $n$ allowed by dipole selection rules. This result for $M_{IF}$ is equivalent to the TPA matrix element $M_{if}$ defined in Eq.\ (\ref{eq:Mif}).
\end{document}